\documentclass[aps,prb,showpacs,twocolumn,groupedaddress]{revtex4}
\usepackage{graphicx}
\usepackage{graphics}
\usepackage{epsfig}
\usepackage{amsmath}
\usepackage{amssymb}
\usepackage{bbm}
\usepackage{verbatim}

\newcommand{\RR}{\ensuremath{\mathbbm{R}}} 
\def\id{\ensuremath{\mathbbm{1}}} 

\newcommand{\Eqref}[1]{Eq.~(\ref{#1})}
\newcommand{\Figref}[1]{Fig.~\ref{#1}}
\def\ket#1{\left|#1\right>}
\def\bra#1{\left<#1\right|}
\newcommand{\ketbra}[2]{\ket{#1}\hspace{-4pt}\bra{#2}}
\newcommand{\proj}[1]{\ketbra{#1}{#1}}

\newcommand{\mr}[1]{\mathrm{#1}}

\newcommand{\up}{\hspace{-1pt}\uparrow}
\newcommand{\dn}{\hspace{-1pt}\downarrow}

\begin{document}

\title{A quantum interface between light and nuclear spins in quantum dots}
\author{Heike Schwager, J. Ignacio Cirac, and G\'eza Giedke}
\affiliation{Max--Planck--Institut f\"{u}r Quantenoptik,
  Hans-Kopfermann--Str. 1, D--85748 Garching, Germany }
\date{\today}

\begin{abstract}
  The coherent coupling of flying photonic qubits to stationary matter-based
  qubits is an essential building block for quantum communication networks. We
  show how such a quantum interface can be realized between a traveling-wave
  optical field and the polarized nuclear spins in a singly charged quantum dot
  strongly coupled to a high-finesse optical cavity. By
  adiabatically eliminating the electron a direct effective coupling is
  achieved. Depending on the laser field applied, interactions that enable
  either write-in or read-out are obtained.
\end{abstract}
\maketitle

\section{Introduction}
The coherent conversion of quantum information between mobile photonic qubits
for communication and stationary material qubits for storage and data
processing is an important building block of quantum networks. In atomic
systems several ideas to realize such a \emph{quantum interface} have been
suggested and experimentally demonstrated in recent years (see \cite{Kim08}
for a review). For semiconductor quantum dots (QD) proposals for interfaces in
analogy to the cavity-based atomic schemes have been put forward \cite{IAB+99}, \cite{YLS05}
and major prerequisites such as strong coupling to a nano-cavity \cite{HBW+07}
have been realized (see \cite{HaAw08} for a review). Here we will show how to
realize a QD-based quantum interface between the \emph{nuclear spins} in a QD
and the optical field. The read-out we propose maps the nuclear state to the
output mode of the cavity directly, while the write-in proceeds by
deterministic creation of entanglement between the nuclear spins and the
cavity output-mode and subsequent teleportation. Our scheme has several
attractive features: the very long nuclear spin lifetimes make the nuclei
attractive for storing quantum information \cite{TIL03} and the use of
collective states makes it possible to map not just qubits but also
multi-photon states. In addition, typical electron spin decoherence processes
will be suppressed: the major such process --hyperfine interaction with the
lattice nuclear spins \cite{SKL03}-- is harnessed to achieve the desired
coupling and the influence of other processes is weakened since the electronic
states can be adiabatically eliminated from the dynamics. The price for this
is a reduction in the speed of the mapping process and the necessity to
initialize the nuclear spin ensemble in a highly polarized state. In view of
the high nuclear polarization of above 80\% reported recently
\cite{Maletinsky2008} the proposed protocol enables the high-fidelity mapping
between a (traveling) optical field and the nuclear spin ensemble in a
realistic setup.

The paper is organized as follows: First, we introduce the system in
Sec.~\ref{sec:system}. In Sec.~\ref{sec:adiabatic} we sketch the
adiabatic elimination that yields the Hamiltonians that describe the
effective coupling between light and nuclear spins (for a detailed
derivation see App.~\ref{app:adiabatic}). Next, we
explain the interface protocol in Sec.~\ref{sec:interface} and
finally give an example for the implementation of the protocol in
Sec.~\ref{sec.impl}.

\begin{figure}[ht]
  \centering
  \includegraphics[scale=0.68]{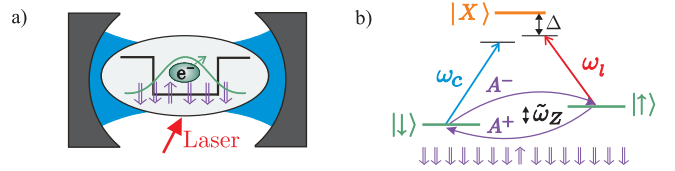}
  \caption{\label{fig:1} (a) Singly charged QD coupled to high-Q
    optical cavity. (b) Level scheme of the QD. Optical and hyperfine
    transitions.}
\end{figure}
\section{System}\label{sec:system}

We consider a self-assembled QD charged with a single
conduction-band electron, whose spin-states
$\ket{\uparrow},\ket{\downarrow}$ are split in a magnetic field.
For clarity we first consider a simplified model, in which both
electronic states are coupled by electric dipole transitions to the
same charged exciton (trion) state $\ket{X}$ in a
$\Lambda$-configuration, cf. Fig.~\ref{fig:1}. Note that the
selection rules in QDs often make it necessary to consider more complicated
level schemes.
After introducing our protocol using this simplified model, we will
present a setting to realize the required coupling and discuss the
effect of corrections to Eq.~(\ref{eq:optical}) in
Sec.~\ref{sec.impl}.

The QD is strongly coupled to a high-Q nano-cavity \cite{HBW+07}.
The two transitions are, respectively, off-resonantly driven by the
cavity mode (frequency $\omega_c$) and a laser of frequency
$\omega_l$, cf. Fig.~\ref{fig:1}, described by the Hamiltonian
\begin{align}\label{eq:optical}
H_\mr{opt}=&\frac{\Omega_c}{2}\,a^{\dagger}\, \ketbra{\downarrow}{X}
+ \frac{\Omega_l}{2}\,e^{+i\omega_l
t}\ketbra{\uparrow}{X}+\textrm{h.c.}\nonumber\\\,&+\omega_c\,
a^{\dagger}a+\omega_{X}\proj{X}+ \omega_z S^z ,
\end{align}
where $\hbar=1$, $\Omega_l, \Omega_c$ are the Rabi frequencies of
laser and cavity fields, $a^{\dagger}$, $a$ are the cavity photons,
$\omega_X$ denotes the trion energy, $\omega_z$ the Zeeman splitting
of the electronic states and
$S^z=1/2(\ketbra{\uparrow}{\uparrow}-\ketbra{\downarrow}{\downarrow})$.
In Sec.~\ref{sec.impl}, we discuss how to effectively realize such a
three-level system in a quantum dot. A detailed discussion of cavity
decay ($\ll\Omega_l, \Omega_c$)
will be considered later on.

As already mentioned, in most QDs the electron spin also has a strong
hyperfine interaction with $N\sim10^4$-$10^6$ lattice nuclear
spins\cite{SKL03}. For s-type electrons it is dominated\footnote{We neglect
  the non-contact parts of the hyperfine interaction \cite{FTCL09} and other
  small nuclear interactions such as the nuclear Zeeman term and the
  interaction between the nuclear spins.} by the Fermi contact
term
\begin{equation}\label{eqn:hfneu}
H_\mathrm{hf}=\frac{A}{2}(S^+A^-+\mr{h.c.}) + A S^zA^z,
\end{equation}
where $A$ is the hyperfine coupling constant, $S^{\pm}$ are the
electron spin operators and $A^{\pm,z}=\sum_j\,\alpha_j I_j^{\pm,z}$
are the collective nuclear spin operators (we consider spin-1/2 nuclei for
simplicity). The individual coupling
constants $\alpha_j$ are proportional to the electron wave function
at site $j$ and normalized to $\sum_j\alpha_j=1$.

A prerequisite for using nuclear spins as a quantum memory is to
initialize them in a highly polarized state which also satisfies
$A^-\ket{\psi_0}=0$, i.e. is decoupled from the electron in state
$\ket{\dn}$ (``dark state''). Recently, nuclear polarization
$P=\left<A^z\right>/(-1/2)$ of $P>80$\% has been reported
\cite{Maletinsky2008} (see also \cite{BSG+05,SNM+08}). The dark
state condition is the natural consequence of using $H_\mr{hf}$ to
polarize the nuclei \cite{IKTZ03}, but has not yet been verified
experimentally. It is useful to separate the large expectation value
of $A^z$, which describes the effective magnetic field experienced
by the electron spin due to the nuclei and write $A^z =
\left<A^z\right>_{\psi_0}+\delta A^z$. Henceforth we include the
first term in $H_\mr{opt}$ by introducing
$\tilde\omega_z = \omega_z+A\langle A^z\rangle_{\psi_0}$.\\
In the high-polarization regime $1-P \ll 1$ a very convenient \textit{bosonic
  description} for
the nuclear spins becomes available: all excitations out of the
fully polarized state and in particular the collective spin operator $A^+$ are
approximated by bosonic creation operators applied to the $N$-mode vacuum
state \cite{Christ2008,KST+Fl09}. Replacing $A^-\to(\sum_j\alpha_j^2)^{1/2}b$
and $A^z\to(-\frac{1}{2}+\frac{1}{N}b^\dag b)$, Eq.~(\ref{eqn:hfneu}) reads
(small corrections omitted in these replacements are discussed in
Appendix~\ref{app:BosonicDesc})
\begin{equation}
  \label{eq:H3}
  \tilde{H}_{\textrm{hf}} = \frac{g_n}{2}(b^{\dagger}S^-+S^+b) +
  \frac{A}{N}S^z\left(b^\dagger b -\frac{N}{2}\right),
\end{equation}
where $g_n = A\sqrt{\sum_j\alpha_j^2}$. The expression
$N_1=(\sum_j\alpha_j^2)^{-1}$ can be seen as the effective number of nuclear
spins to which the electron couples. In the homogeneous case
$\alpha_j=\mr{const}$ we have $N_1=N$. Neglecting very weakly coupled nuclei
we have $N_1\approx N$ and we will just use $N$ in the following.

The bosonic description emphasizes the relation to quantum optical
schemes, gives access to the toolbox for Gaussian states and
operations and allows a more transparent treatment of the
corrections to the ideal Jaynes-Cummings-like coupling of
Eq.~(\ref{eq:H3}); we will make use of this description later on.

\section{Coupling cavity and nuclear spins}\label{sec:adiabatic}

Our aim is to obtain from $H=H_\mr{opt}+H_\mr{hf}$ a direct coupling
between nuclear spins and light. The Hamiltonian $H$ describes a
complicated coupled dynamics of cavity, nuclei and quantum dot.
Instead of making use of the full Hamiltonian (and deriving the
desired mapping, e.g., in the framework of optimal control theory)
we aim for a simpler, more transparent approach. To this end, we
adiabatically eliminate \cite{BPM06} the trion and the electronic
degrees of freedom, which leads to a Hamiltonian $H_{el}$ that
describes a direct coupling between nuclear spins and light. As
explained later, this can be achieved if the couplings (the Rabi
frequency of the laser/cavity, the hyperfine coupling, respectively)
are much weaker than the detunings to the corresponding transition:
\begin{subequations}
  \begin{align}
&\Delta'\gg\Omega_l,\Omega_c\sqrt{n},\label{eq:condelimi}\\
&\sqrt{\Delta'\,\,\tilde{\omega}_z}\gg\Omega_l,\Omega_c\sqrt{n},\label{eq:condelimii}\\
&\tilde{\omega}_z\gg g_n\sqrt{m}.\label{eq:condelimiii}
  \end{align}
\end{subequations}
Here, $\Delta'=\omega_X-\omega_l+\tilde{\omega}_z/2$ is the
detuning, $n$ is the number of cavity photons, and $m$ the number of
nuclear excitations. Note that typically $\tilde\omega_z<\Delta'$
such that condition (\ref{eq:condelimi}) becomes redundant. In
addition to (\ref{eq:condelimi})-(\ref{eq:condelimiii}), we choose
the adjustable parameters such that all first order and second order
processes described by $H$ are off-resonant, but the (third order)
process in which a photon is scattered from the laser into the
cavity while a nuclear spin is flipped down (and its converse) is
resonant. This leads to the desired effective interaction.

The idea of adiabatic elimination is to perturbatively approximate a
given Hamiltonian by removing a subspace from the description that
is populated only with a very low probability due to chosen initial
conditions and detunings or fast decay processes. If initially
unpopulated states (in our case the trion state $\ket{X}$ and the
electronic spin-up state $\ket{\uparrow}$) are only weakly coupled
to the initially occupied states, they remain essentially
unpopulated during the time evolution of the system and can be
eliminated from the description. The higher order transitions via
the eliminated levels appear as additional energy shifts and
couplings in the effective Hamiltonian on the lower-dimensional
subspace.

The starting point is the Hamiltonian $H=H_{\textrm{opt}}+H_{\textrm{hf}}$
given by Eqs.~(\ref{eq:optical}) and (\ref{eqn:hfneu}). In order to get a
time-independent Hamiltonian, we go to a frame rotating with
$U^{\dagger}=\exp{[-i\omega_lt(a^{\dagger}a+ \proj{X})]}$:
\begin{align}\label{rotatingframeham}
H'=&\frac{\Omega_c}{2}(a^{\dagger}\ketbra{\downarrow}{X}
+\text{h.c.})+\frac{\Omega_l}{2}(\ketbra{\uparrow}{X}
+\text{h.c.})+\delta a^{\dagger}a+\tilde{\omega}_zS^z\notag\\&
 +\frac{A}{2}(A^+S^-+S^+A^-)+AS^z\delta
A^z+\Delta\proj{X} ,
\end{align}
with detunings $\Delta=\omega_{X}-\omega_l$ and $\delta=\omega_c-\omega_l$.

Choosing the cavity and laser frequencies, $\omega_c$ and
$\omega_l$, far detuned from the exciton transition and the
splitting of the electronic states $\tilde{\omega}_z$ much larger
than the hyperfine coupling $g_n$, such that conditions
(\ref{eq:condelimi})-(\ref{eq:condelimiii}) are fulfilled, we can
adiabatically eliminate the states $\ket{X}$, $\ket{\uparrow}$. A
detailed derivation of the adiabatic elimination can be found in
Appendix \ref{app:adiabatic}. It yields a Hamiltonian, that
describes an effective coupling between light and nuclear spins
\begin{eqnarray}
  \label{eq:Heff1a}
  H_{el} =&\frac{\Omega_c\Omega_lA}{8\Delta'\tilde{\omega}_z}(aA^+
  +\mr{h.c.})+\omega_1 a^{\dagger}a\notag\\&-\frac{A}{2}  \delta
A^z-\frac{A^2}{4\tilde\omega_z}A^+A^-+T_{nl},
\end{eqnarray}
where the energy of the photons
$\omega_1=\delta-\frac{\Omega_c^2}{4{\Delta'}}$ and the energy of
the nuclear spin excitations $\sim
-\frac{A}{2N}-\frac{A^2}{4N\tilde{\omega}_z}$. By $T_{nl}$ we denote
the nonlinear terms $T_{nl}=\frac{ A^3}{8\tilde{\omega}_z^2}
A^+\delta A^zA^-+\frac{A^2}{4\tilde\omega_{z}^2}\delta
a^{\dagger}aA^+A^-+\frac{\Omega_c^2\delta}{4{\Delta'}^2}a^{\dagger}a^{\dagger}aa$,
which are small
($\|T_{nl}\|\ll\frac{\Omega_c\Omega_lA}{8\Delta'\tilde{\omega}_z} $)
in the situation we consider ($\delta\ll\Omega_c,
g_n/\tilde{\omega_z}\sim\Omega_l/\Delta'\ll1$) and neglected in the
following. In the bosonic description of the nuclear spins that we
introduced in Eq.~(\ref{eq:H3}) the Hamiltonian given by
Eq.~(\ref{eq:Heff1a}) then reads
\begin{eqnarray}
  \label{eq:Heff1b}
  H_{bs} = g (ab^{\dagger}
  +\mr{h.c.})+\omega_1 a^{\dagger}a+\omega_2
  b^{\dagger}b,
\end{eqnarray}
with coupling strength $g$ given by
\begin{equation}\label{eq:gideal}
g=\frac{\Omega_c\Omega_lg_n}{8\Delta'\tilde{\omega}_z}.
\end{equation}
The energy of the nuclear spin excitations can now be written as
$\omega_2=-\frac{A}{2N}-\frac{g_n^2}{4\tilde{\omega}_z}$. For resonant exchange
of excitations between the two systems, we choose $\omega_1=\omega_2$.  Then
$H_\mr{bs}$ describes a beamsplitter-like coupling of the modes $a$ and $b$.
Processes in which absorption (or emission) of a cavity photon is accompanied
by a nuclear spin flip are resonant, and we have thus derived the desired
effective interaction between light and nuclear spins.  Since
$\sqrt{\Omega_c\Omega_l/(\Delta'\tilde\omega_z)}\ll1$ the effective coupling
$g$ is typically $2-3$ orders of magnitude smaller than the hyperfine coupling
$g_n$.

\begin{figure}[ht]
  \centering
  \includegraphics[scale=0.65]{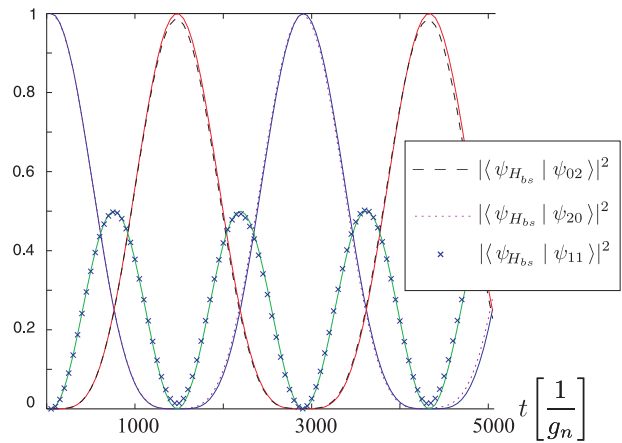}
  \caption{Evolution of the two-photon Fock state $\psi_{20}$ under the full Hamiltonian
  $H'$ (solid lines)
  and Hamiltonian $H_{bs}$ ($\times$, dashed and dotted lines), where the trion and the electronic spin-up state have been eliminated.}
  \label{fig:elimination}
\end{figure}

To illustrate the validity of the adiabatic elimination and the
approximations leading to Eq.~(\ref{eq:Heff1b}), we have simulated
the evolution of the two-photon Fock state $\psi_{20}$ (the first
subscript denotes the number of photons, the second the number of
nuclear spin excitations) under the full Hamiltonian $H'$ given by
Eq.~(\ref{rotatingframeham}) and compared it to the evolution under
the Hamiltonian $H_{bs}$ given by Eq.~(\ref{eq:Heff1b}). We assume
full nuclear spin-down polarization and the validity of the bosonic
description. In the simulation, we choose $\Omega_l=\Omega_c$,
$\Omega_l/\Delta=1/10$, $\Omega_l^2/(\Delta\tilde{\omega}_z)=1/100$
and $g_n/\tilde{\omega}_z=1/50$, such that the conditions given by
Eqs.~(\ref{eq:condelimi})-(\ref{eq:condelimiii}) are fulfilled.
Fig.~\ref{fig:elimination} shows, that $H'$ is well approximated by
$H_{bs}$, and that the nonlinear terms $T_{nl}$ can be neglected.
Almost perfect Rabi-oscillations between the two-photon Fock state
$\psi_{20}$ and the state with two nuclear spin excitations
$\psi_{02}$ can be seen in Fig.~\ref{fig:elimination}. For
$\psi_{01}$, the adiabatic elimination is an even better
approximation to the full Hamiltonian as the nonlinear terms
$T_{nl}$ and the conditions
(\ref{eq:condelimi})-(\ref{eq:condelimiii}) depend on the excitation
number.

In the process leading to the beamsplitter coupling, a photon is
scattered from the  cavity into the laser mode while a nuclear spin
excitation is created (and vice versa). If we interchange the role
of laser and cavity field (i.e., the laser drives the
$\ket{\dn}\leftrightarrow\ket{X}$ transition and the cavity couples
to $\ket{\up}$) then creation of a nuclear spin excitation is
accompanied by scattering of a laser photon \emph{into} the cavity,
i.e. the effective coupling becomes $a^\dag b^\dag+ab$. Tuning the
energies such that $\omega_1=-\omega_2$, the driving laser now
facilitates the \emph{joint} creation (or annihilation) of a spin
excitation and a cavity photon, realizing a two-mode squeezing
effective Hamiltonian
\begin{equation}
  \label{eq:Hsq}
  H_\mr{sq} = g (a^\dagger b^\dagger +a b) + \omega_1 a^\dagger a  + \omega_2
  b^\dagger b.
\end{equation}
Here, the energy of the photons is
$\omega_1=\delta\left(1+\frac{\Omega_c^2}{4{\Delta'}^2}\right)$, the
energy of the nuclear spin excitations is
$\omega_2=-\frac{A}{2N}-\frac{g_n^2}{4\tilde{\omega}_z}$, and the
nonlinear terms are now given by $T_{nl}=\frac{
g_n^2}{4\tilde{\omega}_z^2}\frac{A}{2N}
b^{\dagger}b^{\dagger}bb+\frac{g_n^2}{4\tilde{\omega_{z}}^2}\delta
a^{\dagger}ab^{\dagger}b$. As before, they are much smaller than $g$
and can be neglected for low excitation number. To be able to freely
switch between $H_\mr{bs}$ and $H_\mr{sq}$ simply by turning on and
off the appropriate lasers, both the ``driven'' and the empty mode
should be supported by the cavity.

\section{Quantum Interface}\label{sec:interface}
 Now the obvious route to a quantum interface is
via the Hamiltonian $H_\mr{bs}$: acting for a time $t=\pi/g$ it maps
$a\to ib$ and $b\to ia$ thus realizing (up to a phase) a swap gate
between cavity and nuclear spins. This and related ideas are
explored in \cite{SCG09}. There are two problems with this approach:
Compared to the effective coupling, present-day cavities are ``bad''
with cavity life time $\tau_\mr{cavity}\ll 1/g$, i.e., the cavity
field will decay before its state can be mapped to the nuclei.
Moreover, it is notoriously difficult to couple quantum information
into high-Q cavities, despite proposals \cite{CZKM97} that address
this issue. Both problems can be circumvented for our system by two
key ideas: (i) to include the field modes into which the cavity
decays in the description and (ii) to realize write-in via quantum
teleportation. Moreover, read-out can be realized with similar
techniques. In the following, we assume that all the light leaving
the cavity can be collected and accessed optically. The combination
of strong coupling and high collection efficiency has not yet been
demonstrated for solid-state cavities, although there is remarkable
progress towards that goal \cite{TEFV09}.

Let us first consider the more complicated part, write-in.  In a
first step, the squeezing Hamiltonian $H_\mr{sq}$ (assisted by
cavity decay) generates a strongly entangled two-mode squeezed state
(TMSS) between the nuclear spins and the traveling-wave \emph{output
field} of the cavity. Then quantum teleportation \cite{BrKi98} is
used to deterministically write the state of another traveling-wave
light field onto the nuclear mode.  Similarly, $H_\mr{bs}$ can be used for read-out,
by writing the state of the nuclei to the output field.

Let us now consider $H_{sq}$ and quantitatively derive the entangled
state and discuss the quality of the interface it provides. The
Langevin equation of cavity and nuclear operators is  (for $t\geq0$)
\begin{equation}\label{eq:Langevin}
  \begin{split}
\dot{a}(t) &= -igb(t)^\dag -\frac{\gamma}{2}a-\sqrt{\gamma}c_\mr{in}(t), \\
\dot{b}(t) &= -iga(t)^\dag,
\end{split}
\end{equation}
where we have specialized to the case $\omega_1=-\omega_2$,
transformed to an interaction picture with $H_0=\omega_1(a^\dag a-b^\dag b)$, and performed the
rotating-wave and Markov approximations in the description of the
cavity decay \cite{GZ00}. Here, $c_\mr{in}$ describes the vacuum
noise coupled into the cavity and satisfies
$[c_\mr{in}(t),c_\mr{in}^\dag(t')]=\delta(t-t')$. Integrating Eqs.
(\ref{eq:Langevin}), we get
\begin{equation}\label{eq:sol1}
  \begin{split}
  a(t) &=
\alpha_{1}^-(t)a+\alpha_2(t)b^\dag +
\sqrt{\gamma}\int_0^t\!\!\alpha_{1}^-(t-\tau)c_\mr{in}(\tau)d\tau, \\
  b(t) &=
\alpha_{2}(t)a^\dag+\alpha_{1}^+(t)b+
\sqrt{\gamma}\int^t\!\!\alpha_{2}(t-\tau)c_\mr{in}^\dag(\tau) d\tau,\\
\end{split}
\end{equation}
where $\alpha_{1}^{\pm}(t)=e^{-\gamma t/4}\left[ \cosh(\nu t)\pm
  \gamma/(4\nu)\sinh(\nu t) \right]$, $\alpha_2(t)=-ig/\nu e^{-\gamma
  t/4}\sinh(\nu t)$ and $\nu=\sqrt{(\gamma/4)^2+g^2}$; and $a,b\equiv
a(0),b(0)$ in this equation. It may be remarked here that the analogous
equations with $H_\mr{bs}$ instead of $H_\mr{sq}$
lead to almost identical solutions: now $a(t)$ is coupled to $b(t)$
instead of $b^\dag(t)$ and the only other change to \Eqref{eq:sol1}
is to replace $\nu$ by $\tilde\nu=\sqrt{(\gamma/4)^2-g^2}$.

While \Eqref{eq:sol1} describes a non-unitary time-evolution of the
open cavity-nuclei system, the overall dynamics of system plus
surrounding free field is unitary. It is also Gaussian, since all
involved Hamiltonians are quadratic. Since all initial states are
Gaussian as well the joint state of cavity, nuclei, and output field
is a pure Gaussian state at any time. This simplifies the analysis
of the dynamics and, in particular, the entanglement properties
significantly: For pure states, the entanglement of one subsystem
(e.g., the nuclei) with the rest is given by the entropy of the
reduced state of the subsystem.  Gaussian states are fully
characterized by the first and second moments of the field operators
$R_1 =(a+a^\dag)/\sqrt{2}$ and $R_2 =-i(a-a^\dag)/\sqrt{2}$ via the
covariance matrix (CM) $\Gamma_{kl} = \langle\{R_k,R_l\}\rangle-2\langle
R_k\rangle\langle R_l\rangle$ (where $\{,\}$
 denotes the anticommutator). The CM of the reduced state of a subsystem [e.g.,
$\Gamma_\mr{nuc}(t)$ for the CM of the nuclei at time $t$] is given
by the sub-matrix of $\Gamma$ that refers to covariances of system
operators only. For a single mode, the entropy of the reduced system
can be obtained from the determinant of the reduced CM and with
$x(t)\equiv\det\Gamma_\mr{nuc}(t)$ we get a simple expression for
the entropy (i.e. entanglement):
\begin{equation}
  \label{eq:EoE}
  E(t)=x(t)\log_2x(t)-[x(t)-1]\log_2[x(t)-1].
\end{equation}
Since the state at hand (including the output field) is pure and
Gaussian it is fully determined by $x(t)$ up to local Gaussian
unitaries \cite{GECP03}: it is locally equivalent to a TMSS
$\ket{\psi(r)}=(\cosh r)^{-1}\sum_n(\tanh r)^n\ket{nn}$ with
CM (in $2\times2$ block matrix form)
\[
\Gamma_\mr{TMSS} =
\left(\begin{array}{cc}\cosh(2r)\id_2&\sinh(2r)\sigma_z\\
    \sinh(2r)\sigma_z&\cosh(2r)\id_2 \end{array}\right).
\]
The squeezing parameter $r$ is determined by $x(t)=\cosh^2(2r)$.
From \Eqref{eq:sol1} we find that
$\Gamma_\mr{nuc}(t)=\cosh[2r(t)]\id_2$ for all $t\geq0$, where
$\cosh[r(t)]$ is given by
\begin{eqnarray}\label{eq:coshr}
\cosh r = e^{-\frac{\gamma t}{4}}\left(
\frac{\gamma}{2\nu}\sinh(2\nu
  t)+\frac{g^2+\frac{\gamma^2}{8}}{2\nu^2}\cosh(2\nu t)+\frac{g^2}{2\nu^2} \right)^{1/2}
\end{eqnarray}
and quantifies how strongly the nuclei are entangled with cavity and
output field. After turning off the coupling $g$ at time
$t_\mr{off}$ the nuclei are stationary while the cavity decays to
the vacuum. Therefore, the final entanglement of nuclei and output
field at time $t-t_\mr{off}\gg1/\gamma$ is given by \Eqref{eq:EoE}
with $x(t)=\cosh[2r(t_\mr{off})]^2$. Note that for $\gamma\gg g,1/t$
and keeping only the leading terms in Eq. (\ref{eq:coshr}), $\cosh[2r(t)]$ simplifies to
$3[1-8(g/\gamma)^2]e^{\frac{4g^2}{\gamma}t}$, i.e., two-mode
squeezing $r(t)$ grows linearly with time at rate
$\sim\frac{4g^2}{\gamma}$.

In order to perform the teleportation, a Bell measurement has to be performed
on the output mode of the cavity and the signal state to be teleported. This
is achieved by sending the two states through a 50:50 beam splitter and
measuring the output quadratures \cite{BrKi98}. Hence the output mode of the
cavity, $B_0$, needs to be known to properly match it with the signal mode at
the beam splitter.  It can be expressed as a superposition of the bath
operators $c(x,t)$ as $B_0(t)=\int_\RR z^0(x,t)c(x,t) dx$. By definition, the
mode $B_0$ contains all the photons emitted from the cavity, hence all other
modes $B_{k\not=0}$ (from some complete orthonormal set of modes containing
$B_0$) are in the vacuum state. This implies $\langle B_k(t)
B_l(t)\rangle\propto\delta_{k0}\delta_{l0}$, from which the mode function
$z^0$ can be determined as
\begin{eqnarray}\label{eq:outputmode}
z^0(x,t) &=& \alpha_2(t-x)/\sqrt{\int_{\RR}|\alpha_2(t-x)|^2dx}.
\end{eqnarray}

The procedure for write-in then is: let $H_\mr{sq}$ act for a time
$t_1$ to create the TMSS $\psi(r(t_1))$ of the nuclei entangled with
cavity and output field. To obtain a state in which the nuclei are
only entangled to the output field, we switch the driving laser off
$(g=0)$ and let the cavity decay for a time $t_2\gg \tau_\mr{cav}$,
obtaining an (almost) pure TMSS of the nuclei and the output mode,
which is used for quantum teleportation. Teleportation maps the
state faithfully up to a random displacement $d$, which depends on
the measurement result. This can be undone with the help of
$H_\mr{bs}$ \cite{SCG09} to complete the write-in.

The read-out step follows identical lines, except that $H_\mr{sq}$
is replaced by $H_\mr{bs}$ and no teleportation is necessary since
the state of the nuclei is directly mapped to the output mode of the
cavity; for more details see \cite{SCG09}.

As mentioned, we assume that all light that leaves the cavity can be
collected and further processed. Losses could be modeled by mixing
the outgoing light with yet another vacuum and tracing over the
latter. Considering a fully decayed cavity, the reduced state of
nuclei and output mode is now mixed, but still entangled (unless the
losses are $f=100\%$). Whether or not the state still allows for
better-than classical teleportation depends on $f$ and $r$. E.g.,
for $r=1$ even at losses of 40\%, $F_\mr{tel}>0.7$ (and $>0.5$ even
at 75\% loss). Note, however, that our read-out scheme is much less
tolerant of losses.

\begin{figure}[ht]
  \centering
  \includegraphics[scale=0.65]{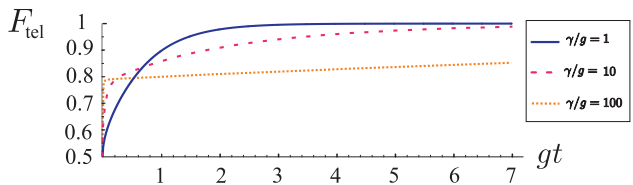}
  \caption{Average fidelity for the mapping of coherent states to the
    nuclei via teleportation (after complete decay of the cavity) plotted as a function of the interaction time $t_\mr{off}$
    for different values of $g/\gamma=1,10,100$ (solid,
    dash-dotted, dashed). All fidelities converge to $1$ as
    $gt\to\infty$. }
  \label{fig:finalent}
\end{figure}

The fidelity with which a quantum state can be teleported onto the nuclei
using the protocol \cite{BrKi98} is a monotonic function of the two-mode
squeezing parameter $r(t_\mr{off})$. A typical benchmark \cite{HWPC05} is the
average fidelity $F$ with which an arbitrary coherent state can be mapped.
For $F\geq2/3$ the quantum channel given by teleportation has a positive
quantum capacity.  If a TMSS is used for teleportation, $F$ has a
simple dependence on the squeezing parameter \cite{Fiu02} and is given by
$F(r) = 1/(1+e^{-2r})$. Thus, if our system parameters $g,\gamma$ and the
interaction time $t=t_\mr{off}$ lead to $\cosh[2r(t_\mr{off})]$ we have an
interface that provides a write-in fidelity $F(r(t_\mr{off}))$, cf.\
\Figref{fig:finalent}.  The fidelity for other subsets of states (including,
e.g., finite dimensional subspaces) can be computed from the coherent state
fidelity \cite{HPC06}.
Already for $r(t_\mr{off})\sim1$ fidelities above $0.8$ are
obtained. As seen from \Figref{fig:finalent} this is achieved for
$gt_\mr{off}\lesssim5$ even for strong decay.  After switching off
the coupling we have to wait for the cavity to decay. Since
typically $\gamma\gg g$ this does not noticeably prolong the
protocol.

\section{Implementation}\label{sec.impl}
Quantum dots generally have a richer level structure than the $\Lambda$ scheme
depicted in \Figref{fig:1}. This and the applicable selection rules imply that
$H_\mr{opt}$ is not exactly realized. In this section we take this into
account and discuss a setting that allows to realize the desired coupling.

\begin{figure}[ht]
  \centering
  \includegraphics[scale=0.65]{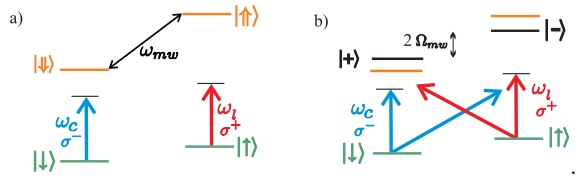}.\caption{ Level scheme of the QD (a) Electronic and trion states split in an external magnetic field in growth direction. They are coupled by a $\sigma^-$-polarized laser and a $\sigma^+$-polarized cavity field with frequencies $\omega_l$, $\omega_c$, respectively. (b) Additional to the setting in (a), a microwave field resonant with the splitting of the trion states in the magnetic field
 ($\omega_{\Uparrow}-\omega_{\Downarrow}=\omega_{mw}$) mixes the trion
 states. Laser and cavity couple both electronic states to the trion states $\ket{+}$ and $\ket{-}$.}
  \label{fig:mixtrion}
\end{figure}

We now consider the two spin states $\ket{\Downarrow},
\ket{\Uparrow}$ of the trion in addition to the two electronic spin
states. We focus on a setup where these states are Zeeman split by
an external magnetic field in growth/$z$-direction (Faraday
geometry). The electronic state $\ket{\uparrow}$ is coupled to
$\ket{\Uparrow}$ (with angular momentum $+3/2$) by $\sigma^+$
circularly polarized light (and $\ket{\downarrow}$ to
$\ket{\Downarrow}$ with $\sigma^-$-polarized light). We can
stimulate these transitions by a $\sigma^-$-polarized cavity field
and a $\sigma^+$-polarized classical laser field, respectively, but
this will not lead to a $\Lambda$  scheme, cf.\
Fig.~\ref{fig:mixtrion}a. The cleanest way to obtain the desired
coupling is to mix the trion states with a resonant microwave field.
The electronic eigenstates are unchanged (being far detuned from the
microwave frequency) and are now both coupled to the new trion
eigenstates $\ket{-}=1/\sqrt{2}(\ket{\Uparrow}-\ket{\Downarrow})$
and $\ket{+}=1/\sqrt{2}(\ket{\Uparrow}+\ket{\Downarrow})$, see
Fig.~\ref{fig:mixtrion}b in a double $\Lambda$ system.
There are other ways to couple both ground states to the same excited state,
e.g., taking advantage of weakened selection rules (due to
heavy-hole/light-hole mixing or an in-plane magnetic field) or using
linearly polarized
light (also in an in-plane magnetic field, i.e. Voigt geometry). They avoid the need of an
additional microwave field at the expense of additional couplings (which have
to be kept off-resonant) and are explored further in \cite{SCG09}.

The Hamiltonian of the system is now given by
\begin{align}\label{eq:opticalim}
H=&\frac{\Omega_c}{2}\,a^{\dagger}\,\ketbra{\downarrow}{\Downarrow}
 +\frac{\Omega_l}{2}\,e^{i\omega_l
t}\ketbra{\uparrow}{\Uparrow}+\Omega_{mw}\,e^{i\omega_{mw}
t}\ketbra{\Downarrow}{\Uparrow}+\textrm{h.c.}\notag\\
&+\omega_c\,a^{\dagger}a+\omega_{\Uparrow}\proj{\Uparrow}+\omega_{\Downarrow}\proj{\Downarrow}+
\tilde{\omega}_z S^z +H_{\text{hf}},
\end{align}
where $\omega_{\Uparrow},\omega_{\Downarrow}=\omega_X\pm\omega_{zh}/2$ include
the hole Zeeman splitting $\omega_{zh}=\omega_{mw}$ and
$H_{\text{hf}}$ is given by Eq.~(\ref{eqn:hfneu}).
In a frame rotating with
\[U^{\dagger}=\exp[{-i(\omega_{mw}+\omega_l)t(\proj{\Uparrow}+a^{\dagger}a) -
  i\omega_l t\proj{\Downarrow}})],\]
the Hamiltonian reads
\begin{align}\label{eq:opticalim2}
H=&\frac{\Omega_c}{2\sqrt{2}}\,(a^{\dagger}\,\ketbra{\downarrow}{+}-a^{\dagger}\,\ketbra{\downarrow}{-})
 + \frac{\Omega_l}{2\sqrt{2}}(\ketbra{\uparrow}{+}+\ketbra{\uparrow}{-})\\\notag
 &+\delta'
a^{\dagger}a+\Delta_+\proj{+}+\Delta_-\proj{-}+ \tilde{\omega}_z S^z
+H_{\text{hf}},
\end{align}
where $\delta'=\omega_c-\omega_l-\omega_{mw}$ and
$\Delta_{\pm}=\omega_{\Downarrow}-\omega_l\pm\Omega_{mw}$. We
adiabatically eliminate $\ket{\pm}$ and $\ket{\uparrow}$ as
explained in Sec.~\ref{sec:adiabatic} and Appendix
\ref{app:adiabatic}.
This yields
\begin{eqnarray}
  \label{eq:Heff1}
  H_{el} = g' (aA^+ +\mr{h.c.}) + \omega_1' a^{\dagger}a - \frac{A}{2}\delta
  A^z - \frac{A^2}{4\tilde\omega_z}A^+A^- + T_{nl}',
\end{eqnarray}
which is of exactly the same form as the Hamiltonian of our toy
model given by  Eq.~(\ref{eq:Heff1a}), and differs only by the
replacements
$\Delta'^{-1}\longrightarrow\frac{1}{2}\left(\Delta_+'^{-1}-\Delta_-'^{-1}\right)$
in the coupling,
$\Delta'^{-1}\longrightarrow\frac{1}{2}\left(\Delta_+'^{-1}+\Delta_-'^{-1}\right)$
in the nuclear energy and
$\Delta'^{-2}\longrightarrow\frac{1}{2}\left(\Delta_+'^{-2}+\Delta_-'^{-2}\right)$
in the nonlinear terms.
As before, the nonlinear terms $T_{nl}'$ are small and are neglected in the
following.  Using the bosonic
description, we then obtain again a beam splitter Hamiltonian
Eq.~(\ref{eq:Heff1b}), where the coupling is now given by
\begin{equation}\label{eq:gimplement}
g'=\frac{\Omega_c\Omega_l
g_n}{16\tilde{\omega}_z}\left(\frac{1}{\Delta_+'}-\frac{1}{\Delta_-'}\right).
\end{equation}
with $\Delta'_{\pm}=\Delta_{\pm}+\frac{\tilde{\omega}_z}{2}$.
Compared to Eq.~(\ref{eq:gideal}) the effective coupling $g$ is
reduced by a factor $\Delta'({\Delta_+'}^{-1}-{\Delta_-'}^{-1})$,
i.e., $\approx2\Omega_{mw}/\Delta'$ for $\Omega_{mw}\ll\Delta'$.

\begin{figure}[ht]
  \centering
  \includegraphics[scale=0.65]{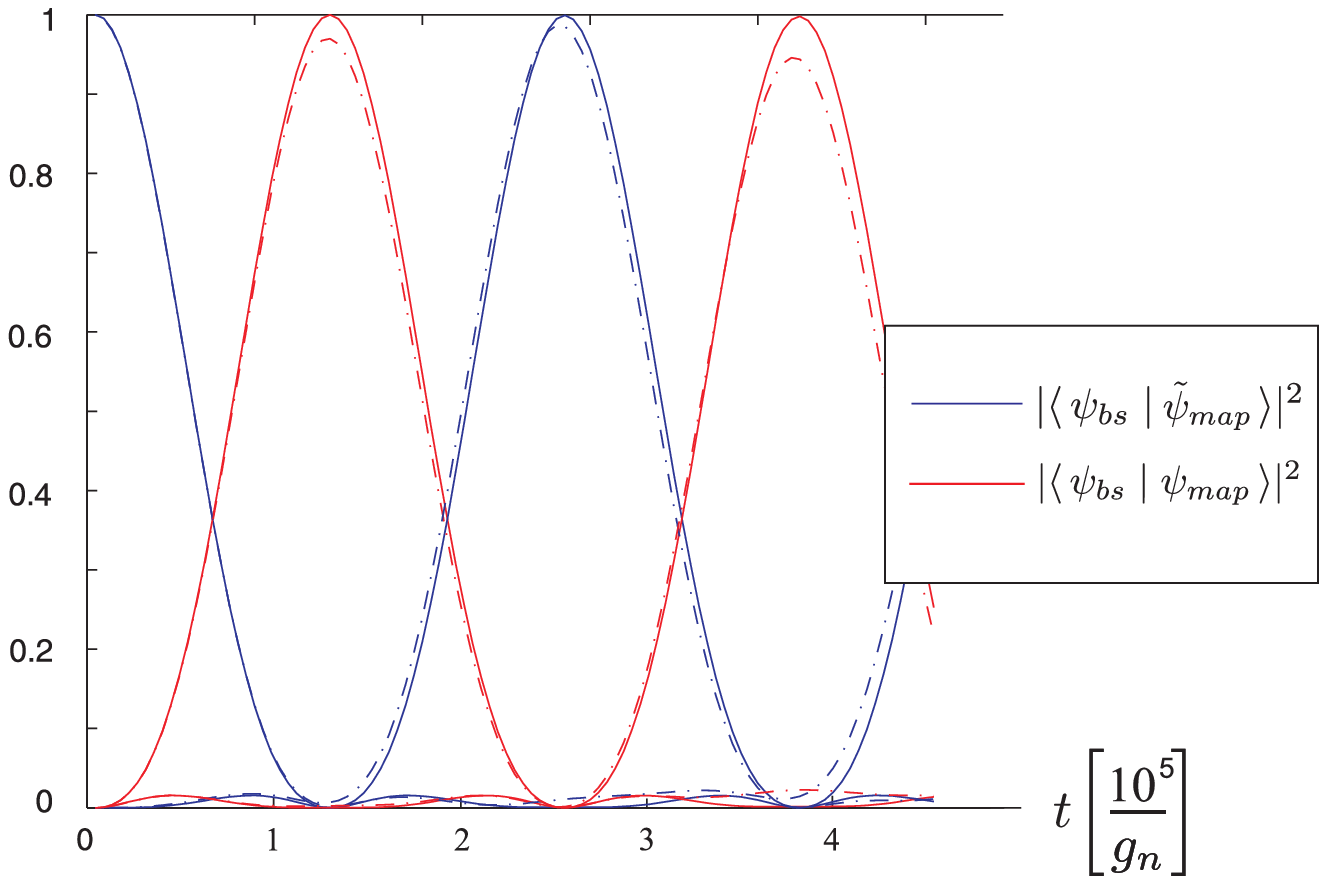}
  \caption{Performance of the quantum interface for the maximally entangled input state $\psi_{in}\propto\sum_{k=1}^2 \ket{k}_R\ket{k}_c$ (subscript $c$ indicates
    the cavity). The red solid curve shows the fidelity $F_{bs}$ of $\psi_{in}$
    evolved under $H_{bs}$ with the ideal target state
$\ket{\psi_{map}}\propto\sum_{k=1}^2
    (-1)^{lk}(i)^k\ket{k}_R\ket{k}_{n}$ (subscript $n$ indicates the nuclei) for
    $gt\,\epsilon[l\pi,(l+1)\pi]$ , where $l$ takes into account the phases acquired during mapping, see text.
The blue solid curve shows the fidelity $\tilde{F}_{bs}$ with
$\ket{\tilde{\psi}_{map}}\propto\sum_{k=1}^2
    (-1)^{lk}\ket{k}_R\ket{k}_c$ for
    $gt\,\epsilon[\frac{2l+1}{2}\pi,\frac{2l+3}{2}\pi]$. Dashed curves depict the same fidelities for evolution under H' (denoted by $F_{H'}/\tilde{F}_{H'}$). (Parameters chosen as in the text).}
  \label{fig:ploteliminationsuperpo}
\end{figure}

To illustrate that $H_{el}$, in the bosonic description, which we
denote by $H_{bs}$, provides a good approximation to $H$ and allows
to implement a good quantum interface, we consider a maximally
entangled state $\sum_k \ket{k}_R\ket{k}_c$ of cavity and some
reference system $R$ and then use the interface to map the state of
the cavity to the nuclei. If a maximally entangled state of $R$ and
nuclei is obtained, it shows that the interface is perfect for the
whole subspace considered. The fidelity of the state $\id_R\otimes
U(t) \sum_{k=1}^2 \ket{k}_R\ket{k}_c\ket{0}_n$ with the maximally
entangled state $\sum_k\ket{k}_R\ket{0}_c\ket{k}_n$ fully quantifies
the quality of the interface. In
Fig.~\ref{fig:ploteliminationsuperpo} we plot this fidelity for the
evolutions $U(t)$ generated by the two Hamiltonians $H$ and $H_{el}$
of Eqs.~(\ref{eq:Heff1}) and (\ref{eq:opticalim2}) to show that a
high-fidelity mapping is possible with the chosen parameters and
that the simple Hamiltonian $H_{el}$ well describes the relevant
dynamics. Since $U(\pi/g)aU(\pi/g)^\dag = ib$ some care must be
taken concerning the phases of the number state basis vectors in the
nuclear spin mode ($\ket{k}_c\mapsto(i)^k\ket{k}_n$) and different
phases at $t=3\pi/g$. For the numerical simulation, we chose the
parameters as follows: the number of nuclei $N=10^4$, the hyperfine
coupling constant $A=100\mu eV$, the laser and cavity Rabi frequency
$\Omega_c=\Omega_l=6\mu eV$, the detuning of the trion
$\omega_{X}-\omega_l=700\mu eV$, the microwave Rabi frequency
$\Omega_{mw}=50\mu eV$ and the effective Zeeman splitting
$\tilde{\omega}_z=50\mu eV$. This corresponds to $\sim4$T using an
electron g-factor of $0.48$ (external and Overhauser field are
counter-aligned) and the corresponding hole Zeeman splitting
$\omega_{mw}\sim 700\mu$eV. With these parameters, a value of
$g\sim5\cdot10^{-5}\mu$eV is obtained, leading to times of $\sim10$
microseconds for an interface operation.

Throughout the discussion we have neglected the internal nuclear dynamics and
corrections to the bosonic description. Nuclear dynamics is caused by direct
dipole-dipole interaction and electron-mediated interaction
\cite{SKL03,YLS06,WiDa06}. In \cite{SCG09} we consider these processes in
detail and show that they are negligible: the coupling of the bosonic mode $b$
to bath modes $b_k$ is by a factor $10^{-2}$ smaller than the coupling $g$ in
$H_\mr{bs}$ given by Eq.~(\ref{eq:gimplement}).\\
The bosonic description of the nuclear spin system can be introduced
in a formally exact way \cite{Christ2008}. However, to obtain the
simple Jaynes-Cummings-like Hamiltonian Eq.~(\ref{eq:Heff1b})
instead of Eq.~(\ref{eq:Heff1a}) we have made several
approximations. As discussed in more detail in Appendix
\ref{app:BosonicDesc}, these can lead to two types of errors, (i) an
inhomogeneous broadening of $\omega_2$ and (ii) leakage from the
mode $b$ due to inhomogeneity. High polarization reduces both
effects. The broadening of $\omega_2$ can be further reduced by an
accurate determination of the Overhauser shift $A^z$. Reduced
Overhauser variance has already been seen experimentally
\cite{Grei07, XU+08, LHZ+Im09}. Leakage is suppressed by the energy
difference of excitations in the mode $b$ and the other modes not
directly coupled to the electron \cite{KST+Fl09} (cf. also the
Appendix).

Finally, sufficiently small electron and cavity decoherence must be
ensured. In particular, we assume the strong coupling limit of
cavity-QED and neglect spontaneous emission for the whole duration
of our protocol, which requires that
$(\Omega_l/\Delta_\pm)^2\gamma_\mr{spont}1/g \ll 1$, where
$\gamma_\mr{spont}$ comprises spontaneous emission of the quantum
dot into non-cavity modes. With the parameters chosen above this
requires $\gamma_\mr{spont}\gg1\mu$s$^{-1}$. Electron spin
relaxation is sufficiently slow in QDs at large Zeeman splitting
($\gtrsim 1$ms) compared to our interaction.  The effect of electron
spin dephasing processes is suppressed by elimination of the
electron: they lead to an inhomogeneous broadening of $g$ and
$\omega_i$ which is small as long as the energy scale of the
dephasing is small compared to the detuning $\tilde\omega_z$.

\section{Conclusion}
We have shown how to realize a quantum interface between the
polarized nuclear spin ensemble in a singly charged quantum dot and
a traveling optical field by engineering beam splitter and two-mode
squeezer Hamiltonians coupling the collective nuclear spin
excitation and the mode of the open cavity. This indicates how to
optically measure and coherently manipulate the nuclear spin state
and opens a path to include nuclear spin memories in quantum
information and communication applications. Moreover, together with
a photo detector for the output mode of the cavity, the quantum
dot--cavity system provides a means to monitor nuclear spin dynamics
on a microsecond time scale and would allow to precisely study the
effect of internal nuclear spin dynamics and the corrections to the
bosonic description used here.

\begin{acknowledgments}
  We acknowledge support by the DFG within SFB 631 and the NIM Cluster of Excellence.
\end{acknowledgments}

\appendix
\section{Adiabatic elimination}\label{app:adiabatic}

In this section, we give a detailed derivation of the adiabatic
elimination that yields the Hamiltonian that describes the effective
interaction between light and nuclei, given by Eq.
(\ref{eq:Heff1a}). The starting point is the Hamiltonian given by
Eq. (\ref{rotatingframeham}).

Choosing the cavity and laser frequencies, $\omega_c$ and
$\omega_l$, far detuned from the exciton transition and the
splitting of the electronic states $\tilde{\omega}_z$ much larger
than the hyperfine coupling $g_n$, such that conditions
(\ref{eq:condelimi}-\ref{eq:condelimiii}) are fulfilled, we can
adiabatically eliminate the states $\ket{X}$, $\ket{\uparrow}$:
denote by $\mathbb{Q} = \ketbra{X}{X}+\ketbra{\uparrow}{\uparrow}$
and
$\mathbb{P}\equiv\openone-\mathbb{Q}=\ketbra{\downarrow}{\downarrow}$
the projectors on the eliminated subspace and its complement,
respectively. Then the Schr\"odinger equation in the two subspaces
reads
\begin{subequations}
\begin{equation}\label{eqn:P1}
E\mathbb{P}\ket{\Psi}=
\mathbb{P}H'(\mathbb{P}+\mathbb{Q})\ket{\Psi},
\end{equation}
\begin{equation}\label{eqn:Q1}
E\mathbb{Q}\ket{\Psi}=\mathbb{Q}H'(\mathbb{P}+\mathbb{Q})\ket{\Psi}.
\end{equation}
\end{subequations}
Our goal is to derive an approximation of the Hamiltonian in the
$\mathbb{P}$-subspace which we denote by $H_{el}$. From
Eq.~(\ref{eqn:Q1}) we obtain
\begin{equation}\label{eqn:Q2}
\mathbb{Q}\ket{\Psi}=
\frac{1}{E-\mathbb{Q}H'\mathbb{Q}}\mathbb{Q}H'\mathbb{P}\ket{\Psi}.
\end{equation}
Inserting Eq.~(\ref{eqn:Q2}) into (\ref{eqn:P1}), we arrive at the
(still exact) equation
\begin{equation}\label{eqn:P2}
E\mathbb{P}\ket{\Psi}=
\left(\mathbb{P}H'\mathbb{P}+\mathbb{P}H'\mathbb{Q}\frac{1}{E-\mathbb{Q}H'\mathbb{Q}}\mathbb{Q}H'\mathbb{P}\right)\mathbb{P}\ket{\Psi},
\end{equation}
for the wavefunction in the electron spin-down subspace, with the
unknown $E$ appearing both on the right hand side (rhs) and the left
hand side (lhs) of Eq. (\ref{eqn:P2}).

Now we use that (i) the range of (unperturbed) energies in the
$\mathbb{P}$-subspace is small compared to the energy difference
between the $\mathbb{P}$- and $\mathbb{Q}$-subspaces and (ii) the
coupling term $\mathbb{P}H'\mathbb{Q}$ is small compared to this
difference, i.e.,
\begin{equation}\label{eq:condel1}
\parallel
\frac{1}{E-\mathbb{Q}H'\mathbb{Q}}\mathbb{Q}H'\mathbb{P}\parallel\ll
1.
\end{equation}
Then the second part on the rhs of Eq.~(\ref{eqn:P2}) is small and
$E$ can be approximated by $E^0$, an eigenvalue of
$\mathbb{P}H'\mathbb{P}=-\left(\frac{\tilde\omega_z}{2}
+\frac{A}{2}\delta A^z-\delta
  a^\dag a \right)\proj{\dn}$, which is here given by
$E^0\approx-\tilde{\omega}_z/2$. Since for our purposes the energy
of the nuclear excitations [$\sim g_n^2/(4\tilde\omega_z)$] and
cavity photons ($\delta$) are chosen equal and are
$\ll\tilde\omega_z$, and $\|\frac{A}{2}\delta A^z\|$ is of order
$\frac{A}{2N}$ and $\ll\tilde\omega_z$, condition (i) is fulfilled.
Condition (ii) given by Eq.~($\ref{eq:condel1}$) is satisfied if the
conditions of Ineq.~(\ref{eq:condelimi}) hold. This yields the
effective Hamiltonian in the electron-spin down subspace:
\begin{equation}
H_{el} =
\left(\mathbb{P}H'\mathbb{P}-\mathbb{P}H'\mathbb{Q}\frac{1}{\tilde\omega_z+\mathbb{Q}H'\mathbb{Q}}\mathbb{Q}H'\mathbb{P}\right)\mathbb{P}.
\end{equation}
To simplify the second term in $H_{el}$  (the denominator is an
operator containing $a,a^{\dagger}, A^-,A^+$), we split it into two
parts: $\tilde\omega_z+\mathbb{Q}H'\mathbb{Q}=B_1+B_2$, where
\begin{equation}
B_1=\tilde{\omega}_z\ketbra{\uparrow}{\uparrow}+(\Delta+\tilde{\omega}_z/2)\proj{X}
\end{equation}
contains the energetically large part and is easy to invert, and
\begin{eqnarray}
B_2=\frac{\Omega_l}{2}(\ketbra{\uparrow}{X}+\mr{h.c.})+\delta
a^{\dagger}a\,\mathbb{Q}
+\frac{A}{2}A^+A^-\ketbra{\uparrow}{\uparrow}.
\end{eqnarray}
contains the Rabi frequency of the laser field $\Omega_l$ that
couples the spin-up state and the trion and the energies of photons
and nuclear spins. From the conditions in Eq.~(\ref{eq:condelimi})
follows that the cavity field is weak and the energies of photons
and nuclear spins are small compared to the energy scale given by
$\Delta'$ and $\tilde{\omega}_z$, therefore
\begin{equation}\label{eq:condel2}
\parallel\frac{1}{\sqrt{B_1}}B_2\frac{1}{\sqrt{B_1}}\parallel\ll 1,
\end{equation} and we can approximate the denominator of Eq.~(\ref{eqn:P2}) by
\begin{equation}\label{eq:b1b2}
\frac{1}{B_1+B_2}\approx\frac{1}{B_1}-\frac{1}{B_1}B_2\frac{1}{B_1}.
\end{equation}
Thus, inserting (\ref{eq:b1b2}) in (\ref{eqn:P2}) and assuming the
conditions given by
Ineqs.~(\ref{eq:condelimi})-(\ref{eq:condelimiii}) to be fulfilled,
we can write the Hamiltonian in the electron spin-down subspace as
\begin{equation}
H_{el}=\mathbb{P}H'\mathbb{P}-\mathbb{P}H'\mathbb{Q}\left(
\frac{1}{B_1}-\frac{1}{B_1}B_2\frac{1}{B_1}\right)\mathbb{Q}
H'\mathbb{P},
\end{equation}
with
$\mathbb{P}H'\mathbb{Q}=\frac{\Omega_c}{2}a^{\dagger}\ketbra{\downarrow}{X}+A
A^+\ketbra{\downarrow}{\uparrow}$, which yields
\begin{eqnarray}
  \label{eq:Heff1a2}
  H_{el} =&\frac{\Omega_c\Omega_lA}{8\Delta'\tilde{\omega}_z}(aA^+
  +\mr{h.c.})+\omega_1 a^{\dagger}a\notag\\&-\frac{A}{2}  \delta
A^z-\frac{A^2}{4\tilde\omega_z}A^+A^-+T_{nl},
\end{eqnarray}
where the energy of the photons
$\omega_1=\delta-\frac{\Omega_c^2}{4{\Delta'}}$ and the energy of
the nuclear spin excitations $\sim
-\frac{A}{2N}-\frac{A^2}{4N\tilde{\omega}_z}$. By $T_{nl}$ we denote
the nonlinear terms $T_{nl}=\frac{ A^3}{8\tilde{\omega}_z^2}
A^+\delta A^zA^-+\frac{A^2}{4\tilde\omega_{z}^2}\delta
a^{\dagger}aA^+A^-+\frac{\Omega_c^2\delta}{4{\Delta'}^2}a^{\dagger}a^{\dagger}aa$,
which are small
($\|T_{nl}\|\ll\frac{\Omega_c\Omega_lA}{8\Delta'\tilde{\omega}_z} $)
in the situation we consider ($\delta\ll\Omega_c,
g_n/\tilde{\omega_z}\sim\Omega_l/\Delta'\ll1$).

\section{Bosonic description of nuclear spins}\label{app:BosonicDesc}
The description of collective spin excitations in a large, highly
polarized system of $N$ spins \footnote{The \emph{relevant} number
$N$ of nuclei coupled to the electron is obtained by neglecting all
very weakly coupled nuclei. For typical choices of the electron wave
function it is of the order of $N_1=\left( \sum_j\alpha_j^2
\right)^{-1}$, which can be determined experimentally by measuring
the variance of $A^z$ in the fully depolarized state.}
$\sigma_j^{\pm,z}$ as bosonic excitations out of the vacuum states
goes back at least to the introduction of the Holstein-Primakoff
transformation \cite{HoPr40}.

If the collective spin operators involved are $A^{\pm,z}\equiv
J^{\pm,z}=\sum_j\sigma^{\pm,z}_j$ and the
system is initialized in the symmetric fully polarized state
$\ket{\dn\dn\dots\dn}$ then the symmetric space spanned by the Dicke states
\cite{ACGT72} $\ket{J=N/2,m}$ is never left under the action
of $A^{\pm,z}$  and up to a $n$-dependent correction the matrix elements
of $J^-$ in the basis $\ket{N/2,n-N/2}$ coincide with the matrix elements of
the bosonic annihilation operator $b$ in the Fock basis $\ket{n}$. In fact we
have
\begin{equation}
  \label{eq:1app}
 \bra{J,n-J}J^-\ket{J,n'-J} =\sqrt{2J}\sqrt{1-\frac{n-1}{2J}}\sqrt{n}\delta_{n,n'-1}.
\end{equation}
As long as $n\ll 2J$ (in the whole subspace significantly populated throughout
the evolution) the factor  $\sum_n\sqrt{1-n/(2J)}P_{\ket{J,n-J}}\approx \id$ and
the association
\begin{subequations}
  \begin{align}
J^+ &\rightarrow \sqrt{2J}b\label{eq:2aapp}\\
\ket{J,n-J}&\rightarrow\ket{n}\label{eq:2bapp}\\
  J^z&\rightarrow -J\id+b^\dag b.\label{eq:2capp}
  \end{align}
\end{subequations}
is accurate to $o(n_\mr{max}/(2J))$. To obtain a more accurate description, we
can even express the factor $\sum_n\sqrt{1-\frac{n-1}{2J}}$ in
Eq,~(\ref{eq:1app}) in bosonic terms, i.e., as $\sqrt{1-b^\dag b/(2J)}$
leading to an \emph{exact} mapping between the spin and bosonic operators.

The intuition we are following is that this association still is useful if we
are dealing with (i) \emph{not fully polarized} systems (i.e., $2J< N$) and
(ii) the collective spin operators appearing in the dynamics are
\emph{inhomogeneous}, i.e. $A^{\pm,z} = \sum_j\alpha_j\sigma^{\pm,z}_j$.

Let us first discuss the two issues separately. If the system is homogeneous
and $J<N/2$ but known, e.g., by measuring $J_z$ and $J^2$, then by
Eq.~(\ref{eq:1app}) compared to the fully polarized case only the parameter
$2J$ has to be adapted and the bosonic description is still good as long as
$n_\mr{max}\ll 2J$.

If $J$ is not precisely known, we get an inhomogeneous broadening of the
coupling constants appearing in front of $A^\pm$ [due to the scaling factor
$\sqrt{2J}$ in Eq.~(\ref{eq:2aapp})] and of the constant in
Eq.~(\ref{eq:2capp}).

If $A^{\pm,z}$ are inhomogeneous, the three operators no longer form a closed
algebra and the dynamics cannot be restricted to the symmetric subspace even
if starting from the fully polarized state. However, it is still possible to
associate $A^-$ to an annihilation operator $A^-\rightarrow
(\sum_j\alpha_j^2)^{1/2}(1+f)b$ where the correction factor $1+f$ is close to
one for highly polarized systems ($\|f\|\sim 1-P$) and depends on the
excitation number not only of the mode $b$ but also of other bosonic modes,
associated with collective spin operators different from $A^\pm$. These can be
introduced, e.g., by choosing a complete orthonormal set of coupling vectors
$\{\vec{\alpha}^{(k)}\}$ with $\alpha^{(0)}\propto\vec{\alpha}$
and defining a
complete set $\{A^\pm_k=\sum_j\alpha^{(k)}_j\sigma^\pm_j, k=0,\dots,N-1\}$ of
collective spin operators. We refer to the modes $b_{k\not=0}$ as ``bath
modes''.

Generalizing the single-mode case discussed before, an exact mapping
$A^-_k\rightarrow (1+f_k) b_k$ and $A^z\rightarrow
-\frac{1}{2}+\frac{1}{N}\sum_k b_k^\dag b_k + C_z$, with operators $f_k,C_z$
describing corrections to the ideal case can be obtained. It was shown in
\cite{Christ2008} that the corrections $f_k, C_z$ are of order $1-P$ for high
polarization. Thus the mapping used in our analysis of the quantum interface
is correct to zeroth order in $1-P$.

Corrections to that description can be obtained by including the corrections
$1-f_k$ and $C_z$. The analysis is simplified by the fact that
coupling between the mode $b$ and the bath modes is weak (first oder in the
small parameter $1-P$) and we are interested only in the mode $b$.
Thus by the replacements \cite{Christ2008}
\begin{subequations}
  \begin{align}
    A^-&\rightarrow (\sum\alpha_j^2)^{1/2}(1-f)b,\\
A^-_k&\rightarrow b_k,\\
A^z&\rightarrow -\frac{1}{2}-\frac{1}{N}\sum_{k=0}^{N-1}b_k^\dag b_k + C_z,
  \end{align}
\end{subequations}
with quadratic hermitian operators $f = \sum_{kk'}\tilde{F}_{kk'}b_k^\dag
b_{k'}$ and $C_z=\sum_{k,k'}C_{kk'}b_k^\dag b_k$ we obtain a first order
description of the dynamics of the mode $b$ (and the electron and photons
coupled to it). Here $C = U\mr{diag}(\alpha_j-1/N)U^\dag$ and $F =
\left(\sum_j\alpha_j^2 \right)U\mr{diag}(\alpha_j^2)U^\dag$ and
and $U$ transforms from the canonical basis to $\left\{ \vec{\alpha}^{(k)}
\right\}$. The matrix $\tilde F$ is obtained from $F$ by multiplying $F_{00}$
by $1/2$ and $F_{k0}, F_{0k}$ by $2/3$. The operators $f,D$ have been chosen
such that the commutation relations of $A^{\pm}$ are preserved to first
order. And while $A^\pm_k,k>0$ are not as accurately preserved, this
affects the dynamics of $A^{\pm,z}$ only to second order \cite{Christ2008}.

From Eq.~(\ref{eq:Heff1a}) we see that that there are three main effects of
the corrections: (i) inhomogeneous broadening of $\tilde\omega_z, g_n$ (and
consequently $\omega_2$ and $g$) due to the finite variance in $P$; (ii)
inhomogeneous broadening of $g$ due to the
variance of the correction factor $1-f$; and (iii) losses of excitations from
the $b$ mode to baths modes due to inhomogeneity.

Since $\tilde\omega_z\gg g_n$, the broadening due to the variance of the
Overhauser field is $\ll\tilde\omega_z$ and thus has only a small effect.
Similarly, the broadening of $g$ affects the form of the output mode $z^0$
[cf. Eq.~(\ref{eq:outputmode})], but since it appears there only via the
parameter $\nu=\sqrt{(\gamma/4)^2\pm g^2}$ the effect is negligible since
$g\ll\gamma$. However, the effective energy of the nuclear excitations,
$\omega_2=g_n^2/(4\tilde\omega_z)$, can be more strongly affected: e.g., a
standard deviation of $10\%$ in $P$ translates to a $10\%$ variation in
$\omega_2$. It must be assured that this variation is small compared to $g$ so
that the resonance condition is maintained.

Concerning leakage, the strongest term is the one arising from $A^z$ and it is
not necessarily small compared to $g$. However, as was pointed out in
\cite{KST+Fl09} the mode $b$ is detuned from the others due to the ``AC Stark
shift'' arising from the off-resonant interaction with the electron [the term
$\sim A^2/(4\tilde\omega_z)A^+A^-$]. As long as this energy shift is large
compared to leakage, losses are suppressed and the mode $b$ is only coupled
dispersively to the bath (via the inhomogeneous broadening). To work in that
regime, $\tilde\omega_z$ must not be too large, i.e., external and Overhauser
field should partially compensate each other while still keeping
$\Omega_l\ll\sqrt{\Delta'\tilde\omega_z}$.


\end{document}